\documentclass[twocolumn,showpacs,preprintnumbers,amsmath,amssymb,prl]{revtex4}

\usepackage{graphicx}
\usepackage{bm}

\begin{document}

\title{Dynamical Crossover in Supercritical Water}

\author{Yu. D. Fomin}
\affiliation{Institute for High Pressure Physics, Russian Academy
of Sciences, Troitsk 142190, Moscow, Russia \\ Moscow Institute of
Physics and Technology, Dolgoprudny, Moscow Region 141700, Russia}
\author{V. N. Ryzhov}
\affiliation{Institute for High Pressure Physics, Russian Academy
of Sciences, Troitsk 142190, Moscow, Russia \\ Moscow Institute of
Physics and Technology, Dolgoprudny, Moscow Region 141700, Russia}
\author{E. N. Tsiok}
\affiliation{Institute for High Pressure Physics, Russian Academy
of Sciences, Troitsk 142190, Moscow, Russia}
\author{V. V. Brazhkin}
\affiliation{Institute for High Pressure Physics, Russian Academy
of Sciences, Troitsk 142190, Moscow, Russia}

\date{\today}

\begin{abstract}
Dynamical crossover in water is studied by means of computer
simulation. The crossover temperature is calculated from the
behavior of velocity autocorrelation functions. The results are
compared with experimental data. It is shown that the qualitative
behavior of the dynamical crossover line is similar to the melting
curve behavior. Importantly, the crossover line belongs to
experimentally achievable $(P,T)$ region which stimulates the
experimental investigation in this field.
\end{abstract}

\pacs{61.20.Gy, 61.20.Ne, 64.60.Kw} \maketitle


In recent decades wide attention of researchers is attracted to
the field of supercritical fluids. Supercritical fluids are of
extreme importance both for fundamental research and for practical
applications especially in chemical industry. In this respect it
is important to give an unambiguous definition of the
supercritical state itself.  For some years a so called Widom
line, which is a line of supercritical maxima of correlation
length and thermodynamic response functions (isobaric heat
capacity $c_P$, isothermal compressibility $k_T$, heat expansion
coefficion $\alpha_P$) in fluids, was used to extend a liquid-gas
coexistence line into supercritical region \cite{widomstanley}.
Later on it was shown that Widom line is ill defined and can not
be used for demarcation of a supercritical region of fluid
\cite{widomvdw,widomlj}. Several authors reported Widom lines of
different systems and arrived to the same conclusion
\cite{widomsw,widomindusy,imre,widomco2}. One can clearly see that
another way of demarcation of gaslike and liquid-like fluids
beyond the critical point was necessary.

Such a way was proposed in our previous publications
\cite{ufn,frpre,frprl}. These publications introduce a so called
Frenkel line which is the line of dynamical crossover in fluids.
It was shown that below the Frenkel line the particles of fluid
make few oscillations at some quasi-equilibrium position following
a jump to another quasi-equilibrium point. This model was proposed
by J. Frenkel \cite{frenkel} after whom the line was named. Above
the Frenkel line the particles of fluid move like in a gas, by
long jumps before a collision with another particle occurs. As a
result below the Frenkel line liquid properties demonstrate some
solidlike behavior whilst above it the properties of liquid are
similar to a dense gas ones. Impact of crossover from crystallike
microscopic dynamics of fluid particles to a gaslike one on
different properties of the fluid was addressed in details in
Refs. \cite{ufn,frpre}. In Ref. \cite{frprl} it was shown that the
most convenient way to find the location of Frenkel line in the
phase diagram is by monitoring a velocity autocorrelation function
(vacf) of the fluid. Basing on this criterion Frenkel line of
several model systems (Lennard-Jones and soft spheres
\cite{ufn,frpre,frprl}) and realistic ones (liquid iron
\cite{frenkel-iron}, carbon dioxide \cite{widomco2,kostyachina},
$TIP4P/2005$ model of water \cite{kostyachina}, methane
\cite{kostyachina} and hydrogen \cite{frenkel-h2}) was calculated.
This work extends the investigations in the field to the most
important liquid - water.

Phase diagram of water is extremely complex. It contains numerous
solid phases including both crystalline and amorphous ones.
However, the fluid part of the phase diagram is no less
interesting. In addition to usual liquid - vapor transition it is
widely assumed that a liquid-liquid phase transition (LLPT) takes
place in unachievable region of (P,T) parameters
\cite{water-llpt}. Water demonstrates a set of liquid state
anomalies such as density anomaly, diffusion anomaly, structural
anomaly and many others \cite{wateranomalies}. Moreover, water has
multiple Widom lines \cite{strong-fr}. The first Widom line is
related to liqud-gas transition while the second one can be
attributed to the hypothetic LLPT. Both Widom lines of water were
vividly discussed in literature (see, for example,
\cite{widom-water-1,widom-water-2,widom-water-3,widom-water-4,imre,vega}
and references therein). In particular, in Ref.
\cite{widom-water-4} a connection between Widom line and dynamical
properties of water was proposed. Since Frenkel line is the line
of dynamical crossover in fluids some kind of relation between the
Widom line and Frenkel line can exist.


In the present work we study the behavior of water by means of
molecular dynamics simulations. An $SPC/E$ model of water is used
\cite{spce}. The phase diagram of this model was reported in
several publications. In \cite{pdcompare} a comparison of solid
part of the phase diagram of $SPC/E$, several variants of $TIP4P$
model and experimental results is given. One can see that all
models fail to reproduce the whole complexity of the experimental
phase diagram, but manage to describe some parts of it. In
particular, $SPC/E$ model is good in describing boiling curve of
water. In Ref. \cite{spce-boiling} boiling curve of $SPC/E$ water
is reported. The critical parameters are found to be $T_c=651.7K$,
$\rho_c=0.326 g/cm^3$ and $P_c=189 bar$. Experimental critical
point of water corresponds to $T_c=647.13K$, $\rho_c=0.322g/cm^3$
and $P_c=220.55 bar$. One can see that except some difference in
critical pressure the critical point of $SPC/E$ model is very
close to the experimental one. Moreover, in Ref. \cite{galli} a
comparison of $SPC/E$ model and ab-initio results at high
pressures and high temperatures was reported. It was shown that
the discrepancy of $SPC/E$ model and ab-initio results is of the
order of  $15-20 \%$ for $T=1000K$ and pressure up to about $100
kbar$. However, at $T=2000K$ and pressures up to approximately
$90kbar$ the agreement of ab-initio and $SPC/E$ results is within
$5 \%$ which should be considered as good agreement. One can guess
that at $T=1000K$ and so high pressure the results are affected by
crystallization effects while at temperatures well above the
melting line $SPC/E$ model can be used to study the high pressure
behavior of water.

A system of $4000$ water molecules in a cubic box was simulated in
molecular dynamics at constant volume, number of particles and
temperature (canonical ensemble). The temperature was held
constant by Nose-Hoover thermostat. The density was varied from
$\rho_{min}=0.8g/cm^3$ up to $\rho_{max}=2.0g/cm^3$ and the
temperatures from $T_{min}=275K$ up to $T_{max}=5 \cdot 10^4 K$.
Initially the system was equilibrated for $1ns$ with a time step
$dt=1fs$. After that it was simulated more $1ps$ with the same
time step in order to calculate the thermodynamic properties.
Finally, $10^5$ steps with timestep $dt=0.1fs$ were made in order
to well reproduce the decay of velocity autocorrelation function
(vacf).

All simulations were performed using lammps simulation package
\cite{lammps}.

\begin{figure}
\includegraphics[width=7cm, height=7cm]{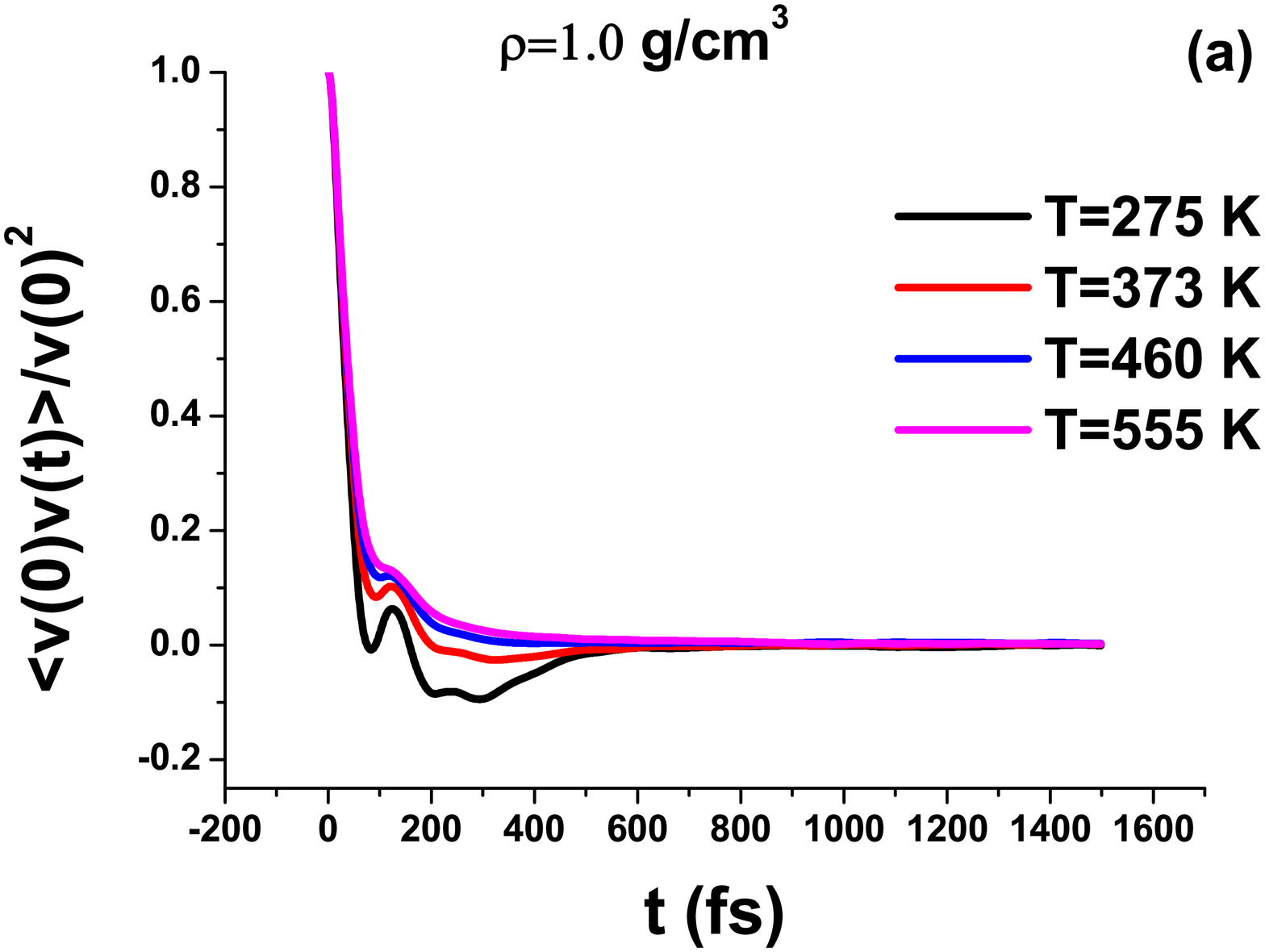}
\includegraphics[width=7cm, height=7cm]{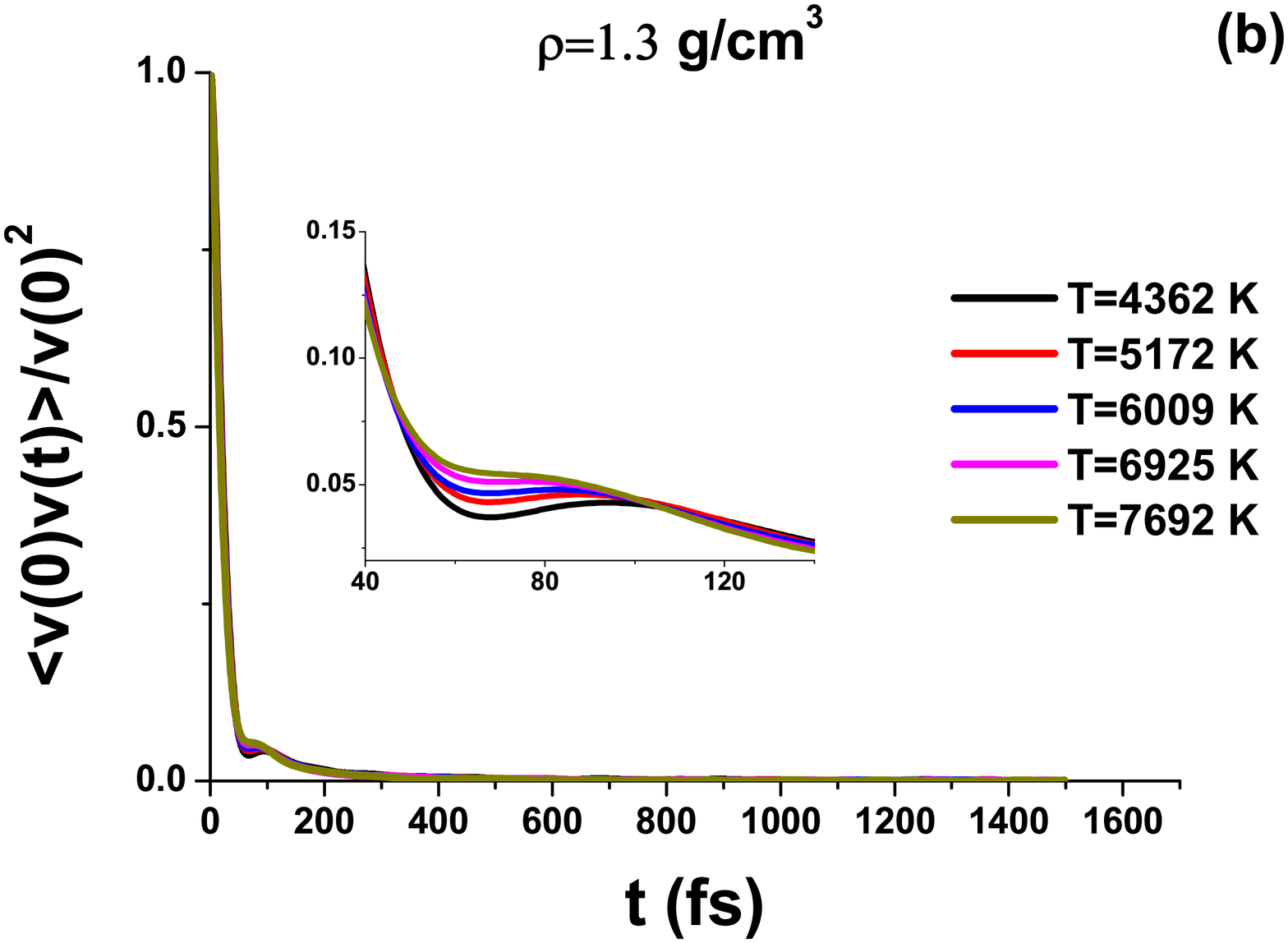}
\caption{\label{fig:fig1} Normalized velocity autocorrelation
functions of oxygen atoms at density (a)$\rho=1.0 g/cm^3$ and (b)
$\rho=1.3 g/cm^3$. The inset in panel (b) enlarges the time period
$40-140$ ps where the oscillation behavior of vacfs takes place.
(Color online).}
\end{figure}

As it was proposed in our earlier publications several methods to
find the location of Frenkel line can be used
\cite{ufn,frpre,frprl}. The most convenient one is based on the
lose of oscillations of vacfs. This criterion is used in the
present work. Figs.~\ref{fig:fig1} (a) and (b) show the vacfs of
oxygens for two densities: $\rho=1.0$ and $1.3$ $g/cm^3$. One can
see that the low temperature vacfs for these two densities look
qualitatively different while at high temperatures they become
very similar. In particular, as the temperature increases the
oscillations of vacfs become less pronounced and finally
disappear.

One more way to estimate the location of Frenkel line in $P-T$ or
$\rho-T$ diagram is related to heat capacity of liquid
\cite{ufn,frpre,frprl}. In case of water the heat capacity $c_V$
undergoes strong decay upon isochoric heating. Next to the melting
line the heat capacity per molecule is about $9k_B$ or $3k_B$ pet
atom ($k_B$ is Boltzmann constant) while at high temperatures it
becomes as small as $1.5k_B$ per atom. The heat capacity per
molecule at Frenkel line should be $6k_B$ per molecule or $2k_B$
per atom. The location of Frenkel line by $c_V$ criterion was
evaluated from experimental data. The data were taken from NIST
database \cite{nist}.

\begin{figure}
\includegraphics[width=6cm, height=6cm]{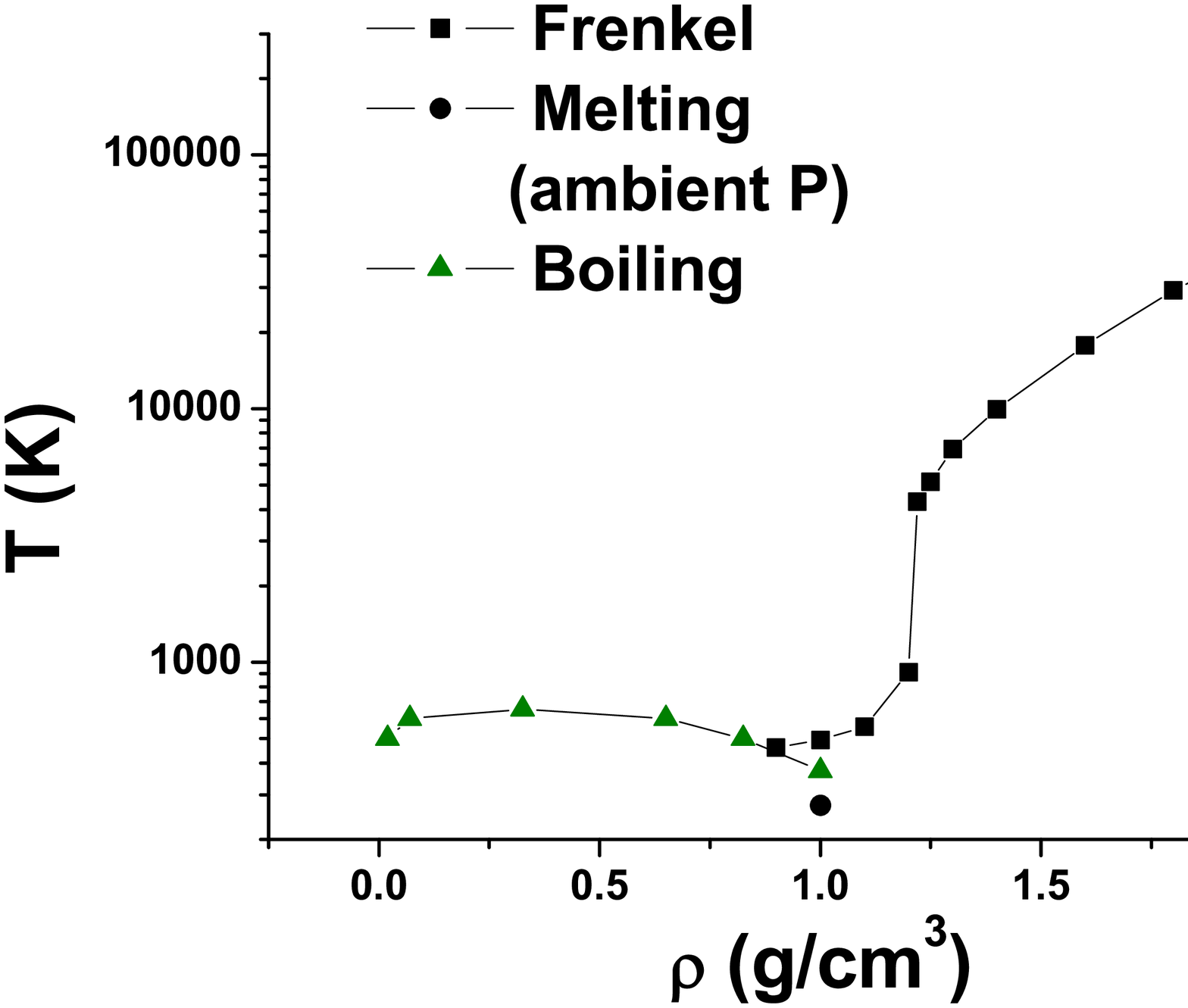}
\includegraphics[width=6cm, height=6cm]{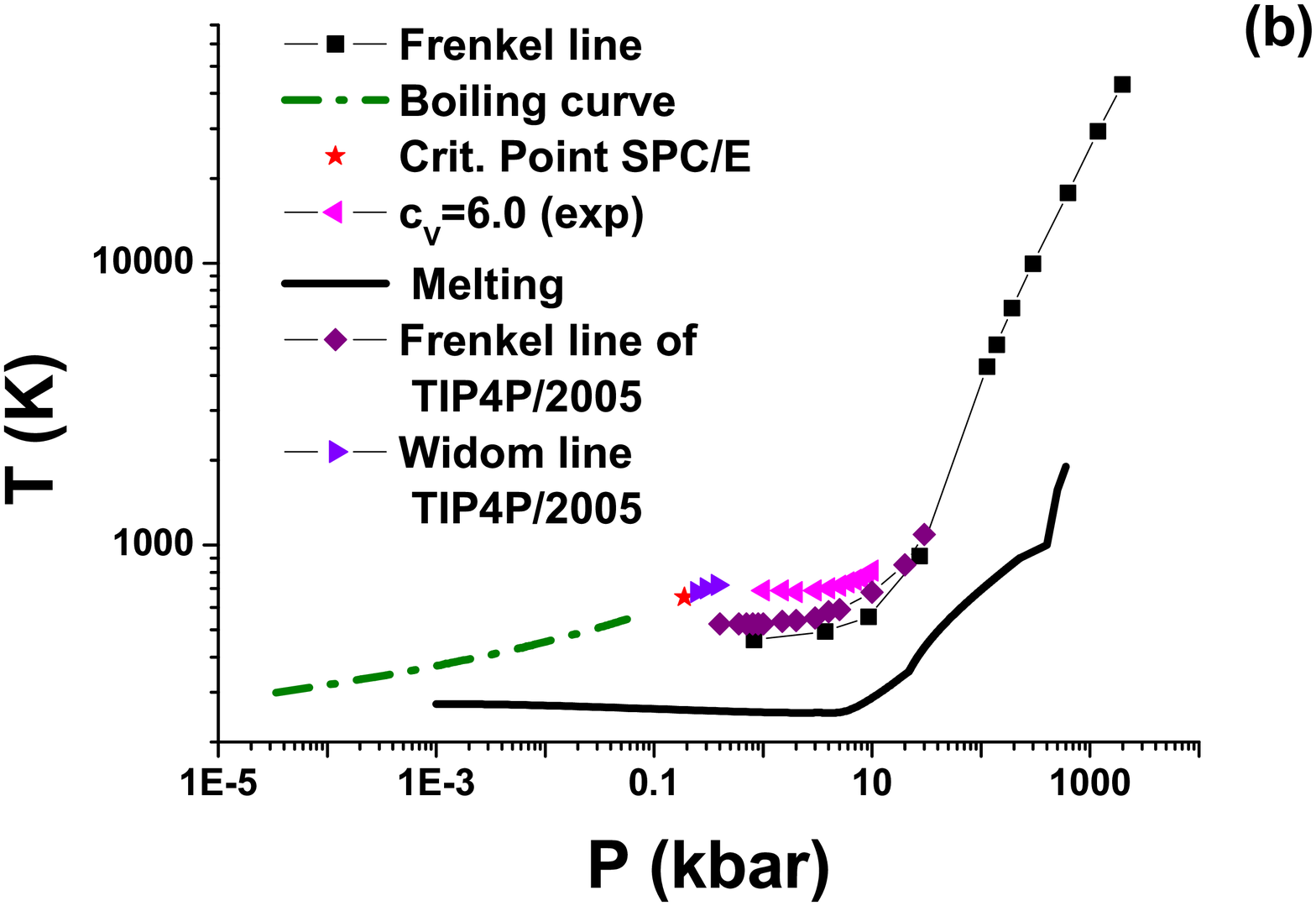}

\includegraphics[width=6cm, height=6cm]{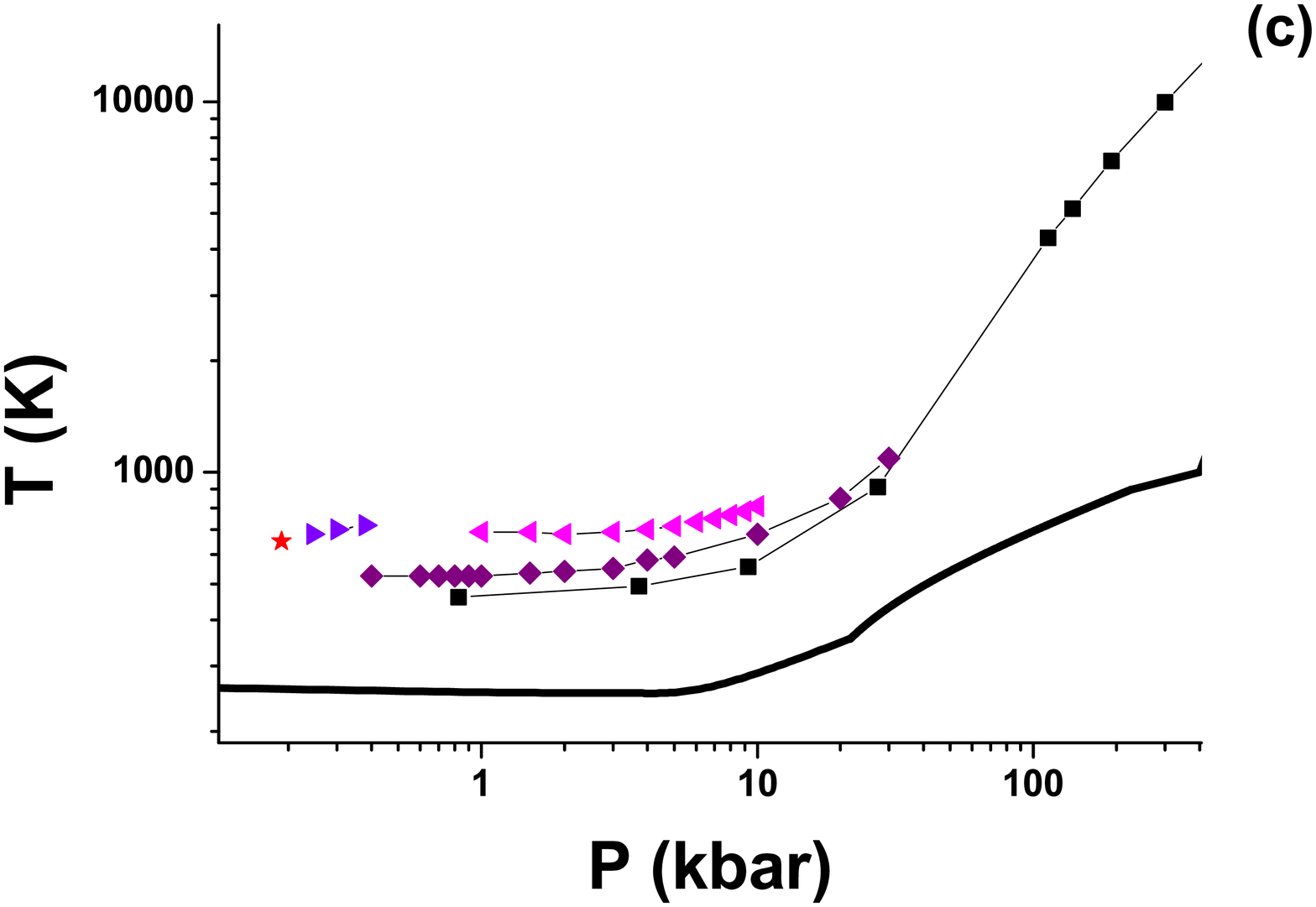}

\caption{\label{fig:fig2} Location of Frenkel line of water in the
phase diagram in (a) $\rho-T$ and (b) $P-T$ coordinates. In case
of $P-T$ diagram the boiling curve corresponds to the experimental
curve \cite{boiling-exp}, the critical point is taken for SPC/E
model \cite{spce-boiling}. The melting line is combined from
several publications
\cite{melt1,melt2,melt3,melt4,melt5,melt6,melt7}. Experimental
data for heat capacity are taken from NIST database \cite{nist}.
The Frenkel line of $TIP4P/2005$ water is taken from Ref.
\cite{kostyachina}. Widom line of $TIP4P/2005$ model is taken from
Ref. \cite{widom-water-4}. Panel (c) enlarges the moderate
pressure part of the diagram. The notation of this panel is the
same that in panel (b). (Color online).}
\end{figure}

Frenkel line of water is shown in Figs.~\ref{fig:fig2} (a) and
(b). From Figs.~\ref{fig:fig2} (a) one can see that Frenkel line
starts at the boiling curve at temperature $T_F=460K \approx 0.7
T_c$, where $T_c$ is the critical temperature. In our previous
publications it was shown that the same ratio $T_F/T_c$ takes
place in Lennard-Jones fluid and liquid iron.

One observes extremely fast grow of the Frenkel line temperature
in a narrow interval of densities $\rho =(1.2-1.22) g/cm^3$. It is
related to extremely slow disappearance of vacf oscillations in
this region.

Figs.~\ref{fig:fig2} (b) shows the location of Frenkel line of
water in $P-T$ phase diagram. Phase diagram of water is very
complex. It demonstrates numerous solid phases. In particular,
usual ice I melting line has a negative slope. Other solid phases
melting lines have positive slopes. The shape of the Frenkel line
of water qualitatively resembles the shape of the melting line. At
low pressures Frenkel line very slowly increases with pressure.
Later on at pressure about $10 kbar$ the slope of the Frenkel line
like the one of the melting curve rapidly increases.

Importantly, the relation between the temperature at the Frenkel
line $T_F$ and the melting temperature $T_m$ changes upon
increasing the pressure. At low pressures (before the rapid
increase) the ratio $T_F/T_m$ is close to $2$. At pressures $P
\approx 100 kbar$ this ratio increases up to $5$. On further rise
of pressure it reaches the value of $9$ at $P \approx 900 kbar$.
It means that in the range of pressures considered in the present
work the Frenkel line bends up with respect to the melting line.

In our previous publications it was proposed that in the limit of
high pressures Frenkel line should be parallel to the melting line
in double logarithmic coordinates. In case of water we are not
aware of any measurements of the melting curve above $1000 kbar$.
We expect that the Frenkel line and the melting curve will be
parallel in the high pressure limit, but one needs to extend the
melting curve to higher pressures in order to check it. Although
this conclusion may be violated by transition of water into
superionic phase under high pressure \cite{melt4}. This phenomena
can not be taken into account in frames of purely classical model
used in the present work.

Very recently Frenkel line of water calculated by vacf criterion
for a different model ($TIP4P/2005$) was reported
\cite{kostyachina}. This line is shown in Fig.~\ref{fig:fig2} (b)
for the sake of comparison. One can see that this line is
systematically higher then our line. However, the lines are very
close to each other and this small discrepancy can be attributed
to the different models under investigation. The authors of
\cite{kostyachina} studied the Frenkel line of water up to
$P_{max}=30kbar$. In our work the Frenkel line is traced up to
pressures as high as almost $P_{max}=2000kbar$ where allowed us to
see an interesting phenomenon. From Figs.~\ref{fig:fig2} (b) one
can see that the bend of Frenkel line in double logarithmic
coordinates takes place at pressure about $50 kbar$. One can
relate this bend to some changes in the local structure of the
liquid.

Figs.~\ref{fig:fig2} (b) and (c) show also the location of Widom
line of $TIP4P/2005$ model of water calculated from maxima of
isobaric heat capacity $c_P$ in Ref. \cite{widom-water-4}.  As it
was shown in several recent publications the supercritical maxima
of different substances rapidly vanish on departing from the
critical point
\cite{imre,widomco2,widomvdw,widomlj,may,widomsw,widomindusy}. In
Ref. \cite{widom-water-4} the Widom line extends up to the
pressure $0.38 kbar$ which is lower then the Frenkel line starts.
Moreover, extrapolation of the Widom line to higher pressures
should go above the Frenkel line of both $SPC/E$ and $TIP4P/2005$
models of water. From this one can conclude that these lines are
not related to each other. One can assume that at low pressure the
crossover of dynamical properties of liquid is governed by the
Widom line while at higher pressures it is determined by the
Frenkel line.

\begin{figure}
\includegraphics[width=7cm, height=7cm]{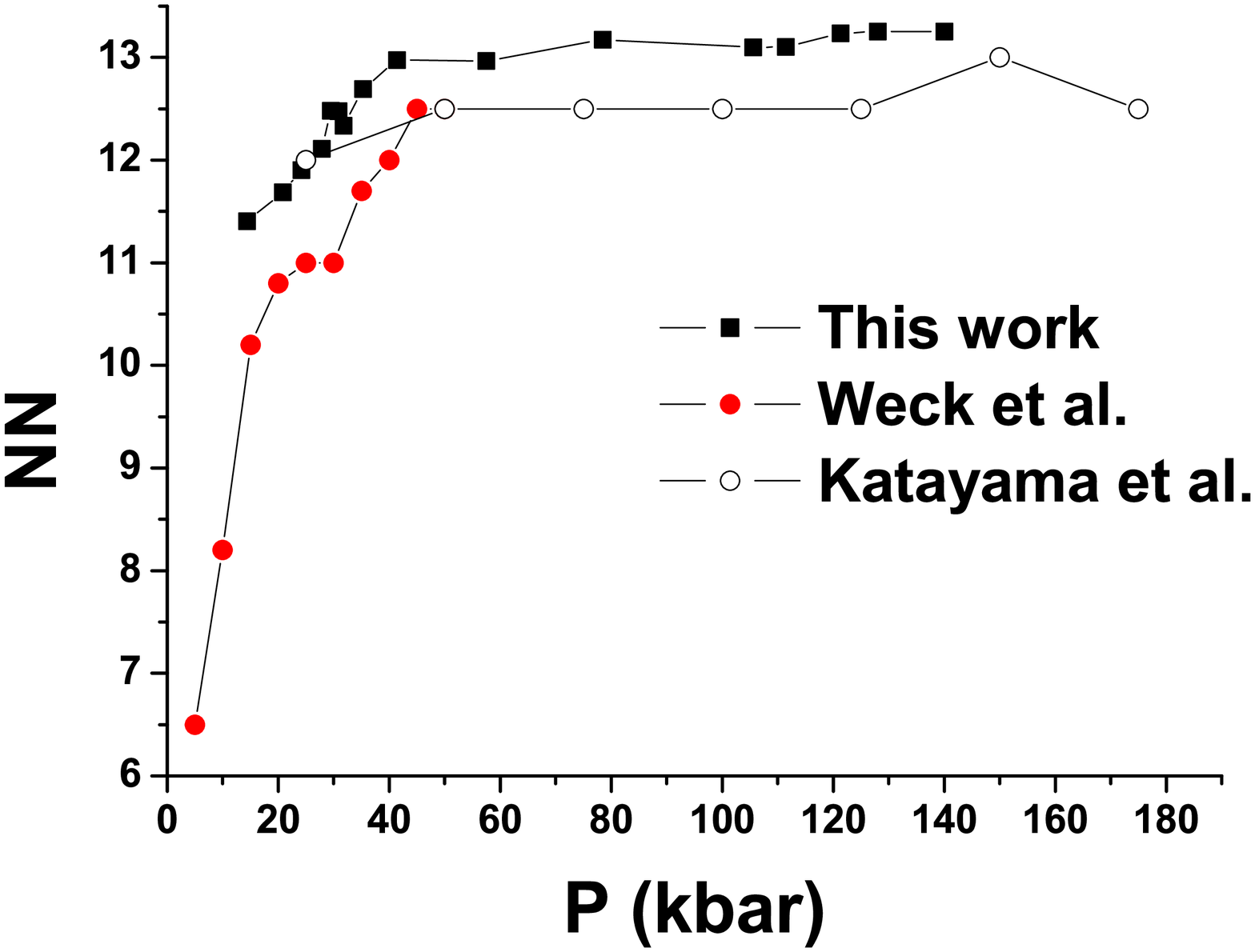}
\caption{\label{fig:fig3} Number of nearest neighbors of oxygen
along $T=1000K$ isotherm. For comparison literature data of NN
along the melting line are also given. The data are taken from
Ref. \cite{nn1} (Weck) and Ref. \cite{nnkatayama} (Katayama).
(Color online).}
\end{figure}

Let us consider the coordination number of oxygens. The
coordination number can be calculated from radial distribution
function (rdf): $NN=4 \pi \rho \int_0^{r_{min}} g(r)r^2dr$, where
$\rho$ is the number density and $r_{min}$ is the location of the
first minimum of rdf.  The number of nearest neighbors (NN) along
the isotherm $T=1000K$ is shown in Fig.~\ref{fig:fig3}. NN of
water along the melting line was reported in Refs.
\cite{nn1,nnkatayama}. Although our results belong to an isotherm
while the results of these publications are related to the melting
line, the NNs are calculated in similar pressure interval. That is
why we show them in the same plot for comparison. The difference
between this work and the literature data should be referred not
only to different lines in $P-T$ plane but also to different
models studied and different methods of calculation. One can see
that up to pressure of approximately $50 kbar$ the coordination
number rapidly increases while above this threshold it holds
approximately constant. One can conclude, that at small pressures
the local structure is very sensitive to the pressure change while
at higher ones the local structure is very stable which is similar
to the case of simple liquid. Therefore one can say that water
becomes "simpler" upon increasing pressure.

We expect that similar rapid increase of Frenkel line in the
density range $\rho= 1.2-1.22 g/cm^3$ can be observed in
$TIP4P/2005$ as well. However, the results reported in the work
\cite{kostyachina} end up at the density $1.2 g/cm^3$. So one
needs to extend them in order to check this assumption.


We report a computational study of dynamical crossover in water.
The temperature of crossover (the Frenkel line temperature $T_F$)
is calculated from the behavior of velocity autocorrelation
functions. The results are compared to the experimental ones
obtained from isochoric heat capacities. It is shown that
qualitative behavior of Frenkel line is similar to the behavior of
the melting curve. However, the ratio $T_F/T_m$ increases with
increasing of pressure. It means that the lines diverge in the
range of pressures considered in the present work. This divergence
can be related to the change of the local structure of water upon
increasing the pressure.

Importantly, the temperatures of dynamical crossover appear to be
rather low at moderate pressures. At pressures $P < 30$ kbar $T_F$
does not exceed $1000$ K. It means that $(P,T)$ parameters of
Frenkel line can be achieved experimentally. Such experimental
works would be important not only for deeper understanding of
dynamical behavior of liquids but also could serve for
supercritical technology. Importantly, the properties of fluids at
the Frenkel line are close to the optimal ones for technological
applications \cite{supercritical}. Water as well as carbon dioxide
are among the most important and widely used supercritical fluids
\cite{book-deb}. Supercritical water is used for many different
applications, such as green solvent, as reaction medium for
different chemical processes, for production of biofuel, oxidation
of hazardous materials which is important for dangerous waste
disposal. Applications of supercritical water include separation,
extraction and purification of different substances and many
others \cite{scw-book}.

The principal property of supercritical water providing its
widespread application is its solving power.  In Ref.
\cite{kostyachina} the solubility of different solutes in carbon
dioxide and its relation to the Frenkel line was discussed. It was
shown that the solubility maxima are close to the Frenkel line.
However, the data for solubility maxima of different solute in
supercritical water are not available at the moment. Basing on the
discussion of \cite{kostyachina} and the results of the present
paper one can propose that the optimal solving power of water
should belong to the interval $T=700 - 1000 K$ and $P=1-3 GPa$.
Currently, the most widespread usage of supercritical water
belongs to the temperatures interval $650-1000K$ and to the
pressures up to $0.5 kbar$. However, these $(P,T)$ conditions were
found empirically and do not have any solid theoretical ground.
Moreover, up to now the supercritical technology advances the
theoretical foundations in the field. This work as well as Ref.
\cite{kostyachina} allow to predict the best $(P,T)$ conditions
for supercritical water application which makes these publications
the pioneering works in developing the theoretical basis of
supercritical technologies.

\bigskip

\begin{acknowledgments}
We are grateful to K. Trachenko for discussions. Yu. F. thanks the
Russian Scientific Center at Kurchatov Institute and Joint
Supercomputing Center of Russian Academy of Science for
computational facilities. The authors are grateful to the Russian
Science Foundation (Grant No 14-22-00093) for the support.
\end{acknowledgments}


\end{document}